%% file: main.tex
\documentclass[10pt,twocolumn,letterpaper]{article}

\usepackage{comment}
\usepackage{multirow}
\usepackage{graphicx}
\usepackage[pagenumbers]{cvpr} 
\input{preamble}

\definecolor{cvprblue}{rgb}{0.21,0.49,0.74}
\usepackage[pagebackref,breaklinks,colorlinks,allcolors=cvprblue]{hyperref}

\def\paperID{} 
\def\confName{CVPR}
\def\confYear{2026}

\title{NOWA: Null-space Optical Watermark for Invisible Capture Fingerprinting and Tamper Localization}
\author{
Edwin Vargas$^{1*}$\quad
Jhon Lopez$^{2}$\quad
Henry Arguello$^{2}$\quad
Ashok Veeraraghavan$^{1}$\\[0.3em]
$^{1}$Rice University\quad
$^{2}$Universidad Industrial de Santander\quad
\\[0.1em]
{\tt\small $^*$edwin.vargas@ieee.org}
\vspace{-1em}
}

\begin{document}
\maketitle

\input{sec/0_abstract}    
\input{sec/1_intro}

\input{sec/2_related}
\input{sec/3_method}

\input{sec/4_results}

\input{sec/5_conclusion}

{
    \small
    \bibliographystyle{ieeenat_fullname}
    \bibliography{main}
}

\input{sec/X_suppl}

\end{document}

%% file: preamble.tex
\definecolor{myorange}{RGB}{255, 165, 0}
\definecolor{mydarkorange}{RGB}{230, 165, 0}
\definecolor{mybrown}{RGB}{155, 78, 20}
\definecolor{mypurple}{RGB}{150, 56, 226}
\definecolor{myblue}{RGB}{10, 50, 230}
\definecolor{myred}{RGB}{230, 10, 10}
\definecolor{mygreen}{RGB}{60, 160, 60}
\definecolor{mydarkgreen}{RGB}{30, 110, 30}
\definecolor{myskyblue}{RGB}{31, 119, 180}
\definecolor{mymagenta}{RGB}{139, 0, 139}
\definecolor{mycyan}{RGB}{0, 100, 100}

\newcommand{\edwin}[1]{{\color{mygreen}{[Edwin: #1]}}}

%% file: sec/0_abstract.tex
%
\begin{abstract}
Ensuring the authenticity and ownership of digital images is increasingly challenging as modern editing tools enable highly realistic forgeries. Existing image protection systems mainly rely on digital watermarking, which is susceptible to sophisticated digital attacks. To address this limitation, we propose a hybrid optical-digital framework that incorporates physical authentication cues during image formation and preserves them through a learned reconstruction process. 
At the optical level, a phase mask in the camera aperture produces a Null-space Optical Watermark (NOWA) that lies in the Null Space of the imaging operator and therefore remains invisible in the captured image. Then, a Null-Space Network (NSN) performs measurement-consistent reconstruction that delivers high-quality protected images while preserving the NOWA signature.
The proposed design enables tamper localization by projecting the image onto the camera's null space and detecting pixel-level inconsistencies. 
Our design preserves perceptual quality, resists common degradations such as compression, and establishes a structural security asymmetry: without access to the optical or NSN parameters, adversaries cannot forge the NOWA signature. 
Experiments with simulations and a prototype camera demonstrate competitive performance in terms of image quality preservation and tamper localization accuracy compared to state-of-the-art digital watermarking and learning-based authentication methods. 
\vspace{-1.em}
\end{abstract}

%% file: sec/1_intro.tex
\section{Introduction}
\label{sec:intro}

We live in an era dominated by digital media, where images and videos have become a primary medium for communication, documentation, and evidence. At the same time, the technological advances that facilitate this ubiquity have also enabled the creation, manipulation, and dissemination of digital images and videos with unprecedented ease, eroding trust in visual media.
From generative editing tools to large-scale diffusion models, synthetic forgeries can now mimic reality with alarming fidelity, prompting the increasingly common question: \textit{Is this real or AI?}.
For instance, a deepfake video of Ukrainian President Volodymyr Zelenskyy in 2022 demonstrates how fake or doctored media artifacts can significantly degrade people’s trust in digital media today. Further, this highlights the urgent need for robust forensic methods for images and videos to detect and counteract media attacks. 

\begin{figure*}[ht!]
\centering
\includegraphics[width=0.88\linewidth]{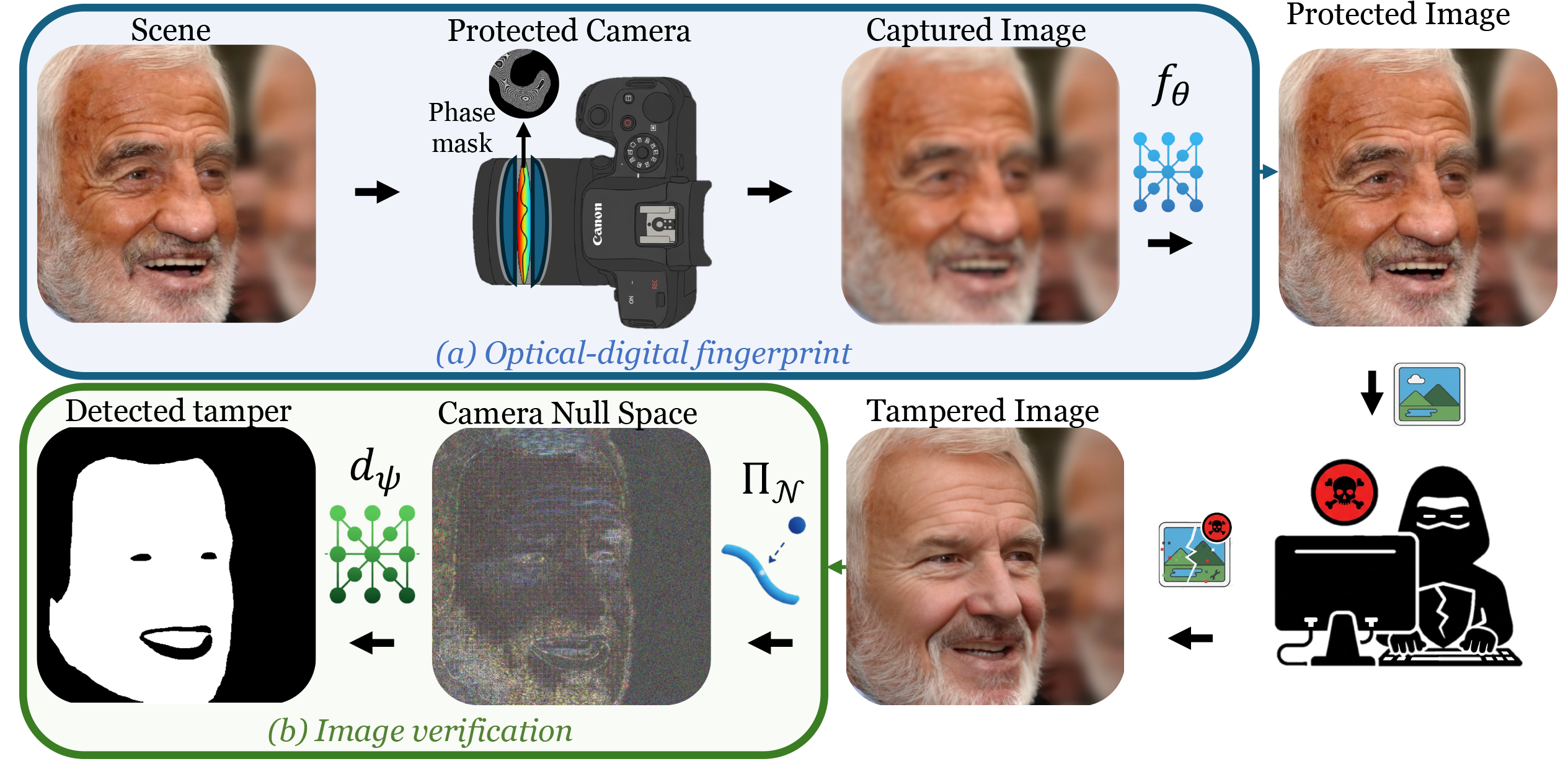}
        \captionof{figure}{\textbf{A hybrid physical-digital pipeline for fragile authentication}.  \textit{(a) Optical-digital fingerprint:} A phase mask (PM) in the camera aperture optically encodes the scene, embedding a unique physical signature before digitization. A neural network $f_\theta$ reconstructs high-quality protected images by recovering the information hidden in the camera’s null space $\mathcal{N}$. ~\textit{(b) Image verification:} The tested image is projected onto $\mathcal{N}$ to obtain a \textit{signature map} revealing the embedded optical fingerprint. A detector $d_\psi$ analyzes this map to detect and localize tampering, by distinguishing predictable system noise from adversarial errors, confirming integrity at the pixel level. 
        }\vspace{-1em}
        \label{fig:teaser}
    \end{figure*}

Most existing protection methods rely on \textit{digital watermarking } \cite{zheng2019survey, farid2009image, kadian2021robust}, which embeds imperceptible codes into images for later verification. 
However, most of these techniques are applied post-capture, making them susceptible to removal or degradation through common editing or compression processes. To overcome this limitation, \textit{optical watermarking} embeds authentication cues at the point of image formation~\cite{jiao2019review, xu2016optical, huang2007optical}, offering hardware-level protection. 
Prior optical watermarking systems have been engineered with a focus on robustness, aiming to preserve watermark integrity under various transformations such as printing, scanning, and lossy compression. However, these systems often rely on complex hardware configurations, including structured light projection and tightly calibrated camera setups \cite{xu2016optical, ma2021optical,ishikawa2013tolerance,michael2025noise}, which can be costly, and difficult to deploy.

While robustness is desirable in copyright protection and media tracing, it can be counterproductive in security tasks that demand sensitivity to even minimal tampering. \textit{Fragile watermarking} addresses this need by making the watermark highly sensitive to changes, supporting pixel-level integrity verification~\cite{zhang2008fragile, singh2017dct, neekhara2024facesigns}. 
However, fragile watermarking remains largely underexplored in the optical domain. 
Recent efforts \cite{arias2025protecting} using coded apertures begin to bridge this gap by embedding imperceptible optical signatures, but exhibit a notably low image quality and only offer a binary authenticity verification without the ability to localize tampering.


Every camera inherently leaves a unique optical signature in its images arising from the physics of its lenses, apertures, and sensors. 
We leverage this intrinsic property for authentication by introducing a custom phase mask (PM) in the optical path of the camera to transform the natural optical signature into a deterministic and verifiable physical-layer watermark embedded during image formation and before the digital life cycle. Unlike post-capture digital watermarks, this optical encoding cannot be removed or accurately reproduced without knowledge of the exact hardware configuration. Additionally, this subtle and spatially distributed modulation by the PM enables the use of compact, standard imaging devices under everyday incoherent illumination, eliminating the need for complex holographic systems or specialized lighting. However, operating solely in raw optical encoded data presents two key challenges: 1)~the encoded images exhibit inherent blur, making them unsuitable for direct visualization or conventional image processing, and 2) access to raw sensor data could allow adversaries to spoof or reverse-engineer the optical signature. These limitations motivate the integration of a learned digital component that restores image quality while preserving the underlying physical signature.

Here, we introduce a \textbf{hybrid optical–digital} framework that unifies the PM encoding with learned digital processing for end-to-end image protection.
This integration allows our system to proactively embed optical authentication cues at the moment of capture while preserving the visual characteristics expected from end-consumer imaging devices. 
More precisely, our approach builds on the observation that the null space of the imaging operator defined by the PM contains information that is optically unmeasurable but mathematically defined. We exploit this property to embed a \textit{Null-space Optical Watermark (NOWA)} that is invisible during capture and recoverable only through the camera’s forward model.
To restore high-quality protected images, we use a Null-Space Network (NSN)~\cite{schwab2019deep}. The NSN learns the inverse optical mapping with measurement consistency, ensuring reconstructions preserve the NOWA.
This design creates a built-in \textit{security asymmetry}: without the optical parameters and trained NSN parameters, the NOWA signature cannot be forged. 

Furthermore, our hybrid design enables flexible and high-accuracy verification via a two-stage mechanism:
isolating the unique optical signature through a null space projection of the designed camera and classifying tampering with a dedicated neural network.
We evaluate our approach using different attack scenarios in simulations, demonstrating competitive tamper localization accuracy compared to state-of-the-art digital watermarking and learning-based authentication methods. We translate our digitally designed phase mask to hardware and demonstrate in real experiments high-quality protected images.

%% file: sec/2_related.tex
\section{Related Work}

\paragraph{Optical Watermarking.}

\textit{Optically Coded Capture.}  
Several works explore embedding authentication cues directly in the optical domain. \citet{ma2022totems} introduce physical refractive objects (\textit{totems}) whose captured regions act as coarse signatures for manipulation detection. Ptychography-based phase encoding enables optical watermarking using single-shot and non-mechanical setups \cite{xu2016optical,ma2021optical}. Natural chromatic aberrations can also serve as intrinsic encoders \cite{mayer2018accurate}, revealing forged regions through aberration inconsistencies. Sensor-level fingerprints have been explored via pattern noise \cite{lukavs2006detecting} and learned noise residuals \cite{cozzolino2019noiseprint}, enabling device-specific forgery localization.

\textit{Coded Illumination.}  
Watermarks can also be embedded via structured lighting. Controlled illumination schemes \cite{tan2009optical,ishikawa2010practical,ishikawa2013tolerance} and noise-coded illumination \cite{michael2025noise} enable spatial and temporal forgery detection. Related works extend this to single-pixel imaging \cite{ye2019image}, display-based encoding \cite{gerstner2022detecting}, and time-varying illumination for simultaneous imaging and concealment \cite{ye2020simultaneous}.

\textit{Optical Coding for Encryption.}  
Optical encryption methods \cite{refregier1995optical} and hardware-based key systems such as OpEnCam \cite{khan2024opencam} encode scenes using proprietary masks, ensuring decryption only with known hardware keys. Similarly, color-coded apertures \cite{arias2025protecting} embed authentication signatures during capture, though at the cost of degraded image quality.
\vspace{-1em}
\paragraph{Digital Watermarking.}
Deep learning has transformed digital watermarking by improving both imperceptibility and robustness. HiDDeN \cite{zhu2018hidden} first introduced an end-to-end encoder–decoder for data embedding and recovery, followed by flow-based generative models \cite{fang2023flow,ma2022towards} that exploit invertibility for high-fidelity watermark extraction. Differentiable augmentations during training, such as JPEG compression and realistic perturbations, enhance robustness to signal distortions \cite{ahmadi2020redmark,fang2022pimog,liu2019novel,wu2023sepmark}. Self-supervised embedding further improves watermark security \cite{fernandez2022watermarking}.
Recent efforts focus on watermark resilience under AI-generated content (AIGC). Robust-Wide \cite{hu2024robust} and VINE \cite{lu2024robust} use diffusion models \cite{sauer2024adversarial} for robust watermark recovery under text-guided edits, albeit with high computational cost. Proactive watermarking introduces identity-sensitive and tamper-aware encoding \cite{zhao2023proactive}, shifting from passive protection to active defense. EditGuard \cite{zhang2024editguard} extends this idea by unifying copyright verification with tamper localization through dual-invisible watermarks, while OmniGuard \cite{zhang2025omniguard} generalizes the approach to high-resolution, manipulation-aware scenarios.

\vspace{-1em}
\paragraph{Computational Photography.}
The co-design of optics and algorithms is at the core of computational photography. This paradigm has been vastly accelerated by modern automatic differentiation programming tools \cite{paszke2017automatic,abadi2016tensorflow}, which enable the implementation of fully differentiable pipelines. This approach has driven innovation across numerous applications, including color imaging and demosaicing \cite{chakrabarti2016learning}, extended depth of field \cite{sitzmann2018end}, depth imaging \cite{chang2019deep,wu2019phasecam3d, lopez2024depth}, spectral imaging \cite{vargas2021time, arguello2023deep, vargas2023learning}, high dynamic range (HDR) imaging \cite{metzler2019deep}, image classification \cite{chang2018hybrid, muthumbi2019learned}, microscopy \cite{jin2024deepdof} and many others. Our work extends this end-to-end optimization technique for the purpose of authentication.

%% file: sec/3_method.tex


\section{Method}
We propose an authentication method that leverages the advantages of a physical and digital solution.
Our proposed hybrid authentication system integrates optical front-end engineering with a Null-Space Network to form a differentiable end-to-end pipeline. The light propagation through the optical system is simulated using Fourier optics \cite{goodman2005fourier, kellman2019phase, wu2019phasecam3d}, enabling the joint optimization of both the optical parameters and the neural network for the authentication task. 
Mathematically, the proposed hybrid watermarked image is
\begin{equation}
 \mathbf{x}_{p} = f_\theta(g_\phi(\mathbf{x})), 
\end{equation}
where $\mathbf{x} \in \mathbb{R}^n$ denotes the input scene, $g_\phi(\cdot)$ models the optical image formation process using a phase mask parameterized by $\phi$, and $f_\theta(\cdot)$ is the reconstruction NSN with parameters $\theta$, responsible for processing the optical encoded images and finally obtain the protected image $\mathbf{x}_p \in \mathbb{R}^n$. 
Following reconstruction, the final stage of our pipeline is a verification mechanism that operates on the protected image $\mathbf{x}_p$ to detect and localize tampering at the pixel level.
In the following, we present more details about the imaging model, NSN, and tamper detection and localization.

\subsection{{Preliminaries}}
\paragraph{{Imaging Model.}}
Our optical system $g_\phi$ consists of a conventional imaging lens and a custom phase mask at the pupil plane. 
The combined effect of the lens and the phase mask defines the system’s point spread function (PSF), which encodes 
a unique physical signature into every captured image. In the paraxial regime, the corresponding phase shift imparted by the convex lens  thickness profile with focal length $f$ and wavelength-dependent refractive index $n(\lambda)$ produces the lens transmission function \cite{goodman2005introduction}

\begin{equation}
    t_L(x, y) = \exp\left(- i \frac{k}{2f} (x^2 + y^2)\right),
    \label{eq:lens_tf}
\end{equation}
with $k = 2\pi / \lambda$ denoting the wavenumber and $(x,y)$ the spatial coordinates on the normalized pupil plane.
To model the additional modulation introduced by the engineered phase mask, we define a spatially varying height profile 
$h_\phi(x, y)$ that produces a wavelength-dependent phase delay
\begin{equation}
    \phi_M(x, y) = k \left(n_M(\lambda) - 1\right) h_\phi(x, y),
    \label{eq:mask_phase}
\end{equation}
where $n_M(\lambda)$ is the refractive index of the mask material. 
The corresponding transmission function of the mask is then
\begin{equation}
    t_M(x, y) = \exp\left( i \, \phi_M(x, y) \right).
    \label{eq:mask_tf}
\end{equation}
The combined pupil function of the lens-phase mask optical system is given by
\begin{equation}
    P(x, y) = A(x, y) \, t_L(x, y) \, t_M(x, y),
    \label{eq:pupil_function}
\end{equation}
where $A(x, y)$ is an aperture function that equals $1$ within the open pupil and $0$ elsewhere.
Given an input optical field $U_{\text{in}}(x, y)$ incident on the pupil, the modulated field immediately after the lens–mask assembly is
\begin{equation}
    U_{\text{out}}(x, y) = P(x, y) \, U_{\text{in}}(x, y).
    \label{eq:field_after_pupil}
\end{equation}
This field then propagates a distance $s$ to the sensor plane according to the angular spectrum propagation model \cite{goodman2005introduction} resulting in
\begin{equation}
    U_{\text{sensor}}(x', y') = 
    \mathcal{F}^{-1}\left\{
    \mathcal{F}\{U_{\text{out}}(x, y)\} \cdot H_s(f_x, f_y)
    \right\},
    \label{eq:propagation}
\end{equation}
$$
    \textnormal{with} \hspace{1em} H_s(f_x, f_y) = \exp\left[i k s \sqrt{1 - (\lambda f_x)^2 - (\lambda f_y)^2}\right],
    \label{eq:transfer_function}
$$
$(f_x, f_y)$ denote spatial frequency coordinates, $\mathcal{F}\{\cdot\}$ and $\mathcal{F}^{-1}\{\cdot\}$ denote the forward and inverse 2D Fourier transforms, and $x',y'$ denote the spatial coordinates on the sensor plane. Finally, the sensor measures light intensity, and the wavelength-specific PSF is given by
\begin{equation}
    p_{\phi,\lambda}(x', y') = 
    \left| U_{\text{sensor}}(x', y') \right|^2.
    \label{eq:psf}
\end{equation}
%
%
Given this formulation, the sensor measurement in the color channel centered at wavelength $\lambda$, i.e., $\mathbf{y}_\lambda \in \mathbb{R}^n$, can be approximated as the convolution between the input scene $\mathbf{x}_\lambda$ and the engineered PSF
\begin{equation}
\mathbf{y}_\lambda = g_{\phi}(\mathbf{x}_\lambda) = \mathbf{x}_\lambda * \mathbf{p}_{\phi,\lambda} + \mathbf{n},
\label{eq:forward_model_conv}
\end{equation}
where $*$ is the convolution operator, $\mathbf{p}_{\phi,\lambda}$ is the vector representation of the the PSF, and $\mathbf{n} \in \mathbb{R}^n$ models additive noise. The design parameters $\phi$ directly control the optical response through $h_\phi(x,y)$, enabling programmable modulation of the image formation process. This Fourier optics formulation follows the standard treatment in computational imaging and phase mask design~\cite{goodman2005fourier, kellman2019phase, wu2019phasecam3d}. For clarity, we henceforth omit the wavelength subscript and assume each color channel is processed independently.

\paragraph{PSF Parameterization.} To ensure smoothness and manufacturability of the designed phase mask, we parameterize the height profile $h_\phi(u,v)$ using a truncated expansion of \emph{Zernike polynomials}:
\begin{equation}
h_\phi(x,y) = \sum_{k=1}^{K} \phi_k Z_k(x,y),
\end{equation}
where $Z_k(x,y)$ denotes the $k$-th Zernike mode and $\phi_k$ its corresponding coefficient. 
This parameterization constrains the optimization space to physically realizable surfaces while maintaining differentiability for end-to-end learning. The coefficients $\phi_k$ are optimized jointly with the digital reconstruction network through gradient-based optimization, allowing the optical and computational components to co-adapt for optimal image authentication performance. 


\color{black}
\paragraph{Null-Space Networks for Reconstruction. }

The measurement process in~\eqref{eq:forward_model_conv} can be modeled as  
\vspace{-0.5em}
\begin{equation}
    \mathbf{y} = \mathbf{A}_\phi \mathbf{x} + \mathbf{n},
\end{equation}
where $\mathbf{A}_\phi$ encodes the optical forward model determined by phase mask and aperture parameters $\phi$.  
Due to diffraction and sampling limits, $\mathbf{A}_\phi$ is ill-conditioned, meaning that only part of the signal is observable, while information in its null space,
\vspace{-0.5em}
\small{
\begin{equation}  
\mathcal{N}(\mathbf{A}_\phi)=\{\,\mathbf{z}\mid \mathbf{A}_\phi\mathbf{z}=0\,\},
\end{equation}}
\normalsize
is completely invisible to the sensor. To address this, Null-Space Networks (NSNs)~\cite{schaub2021nullspace} have been designed to both reconstruct an accurate signal and the lost information in $\mathcal{N}(\mathbf{A}_\phi)$. 
More precisely, given a regularized estimate $\hat{\mathbf{x}}_r=r(\mathbf{y})\in\mathcal{N}(\mathbf{A}_\phi)^\perp$, obtained via a regularized inverse (e.g., pseudoinverse, Tikhonov, or Wiener deconvolution), the NSN recovers an image defined as
\vspace{-0.5em}
\begin{equation}
    \mathbf{x}_p = f_\theta(\hat{\mathbf{x}}_r)
    = \hat{\mathbf{x}}_r
    + \Pi_\mathcal{N}\,
    \mathcal{U}_\theta(\hat{\mathbf{x}}_r),
    \label{eq:nullspace}
\end{equation}
where $\Pi_\mathcal{N}=\mathbf{I}-\mathbf{A}_\phi^\dagger\mathbf{A}_\phi$ is the orthogonal projector onto $\mathcal{N}$.  
This formulation enforces strict \textit{measurement consistency},  
$\mathbf{A}_\phi f_\theta(\hat{\mathbf{x}}_r)=\mathbf{A}_\phi\hat{\mathbf{x}}_r=\mathbf{y}$,  
while allowing the learned network $\mathcal{U}_\theta$ to recover the unobservable components.

\subsection{Null-space Optical Watermark (NOWA)}
We propose to exploit the unobservable subspace $\mathcal{N}(\mathbf{A}_\phi)$ as a secure channel to embed a \textit{Null-space Optical Watermark (NOWA)} that cannot be imitated without access to $\mathbf{A}_\phi$. Unlike digital watermarks, the NOWA is physically encoded during image formation and remains invisible in the raw capture.
However, the encoded images are not directly usable for final users, and any standard reconstruction algorithm would treat the NOWA as noise and remove it. To prevent this, we use the NSN framework to recover the protected image $\mathbf{x}_p$ while anchoring the NOWA within it. 
By leveraging \textit{measurement consistency}, the null space provides a mathematically secure embedding: its energy is annihilated by $\mathbf{A}_\phi$ but recoverable through projection $\Pi_\mathcal{N}$.   
Any modification to $\mathbf{x}_p$ that is inconsistent with $\mathcal{N}(\mathbf{A}_\phi)$ can be detected by re-projecting the image and analyzing the residual null-space statistics. 
This observation forms the foundation of our detection and localization pipeline.
\vspace{-1em}
\color{black}
\paragraph{Tamper Detection and Localization.}
The authentication stage of our framework operates entirely in the null-space domain defined by $\mathbf{A}_\phi$. 
Given a protected reconstruction $\mathbf{x}_p$, we extract a raw \textit{signature map} by projecting it onto $\mathcal{N}(\mathbf{A}_\phi)$ 
\begin{equation}
    \mathbf{s} = \Pi_{\mathcal{N}}(\mathbf{x}_p) = \Pi_{\mathcal{N}}\, \mathcal{U}_\theta(\hat{\mathbf{x}}_r),
    \label{eq:signature_map}
\end{equation}
where we use that $\Pi_{\mathcal{N}}(\hat{\mathbf{x}}_r)=\mathbf{0}$ for 
$\hat{\mathbf{x}}_r \in \mathcal{N}(\mathbf{A}_\phi)^\perp$ and $\Pi_{\mathcal{N}}^2=\Pi_{\mathcal{N}}$, thus isolating the learned component constrained to the unobservable subspace of the optical model. 
This representation captures physically consistent but measurement-invisible features that 
serve as the intrinsic signature of the designed optics.
For genuine images, $\mathbf{s}$ exhibits a stable, predictable spatial pattern determined by the optical encoding. 
Tampered regions, in contrast, disrupt this structure, leading to anomalous null-space responses. 
To interpret anomaly deviations in the null-space, we employ a convolutional neural network (CNN) detector $d_\psi$ that maps $\mathbf{s}$ to per-pixel authenticity probabilities:
\begin{equation}
    \mathbf{m} = d_\psi(\mathbf{s}),
    \label{eq:detector_output}
\end{equation}
where $\mathbf{m}_i \in [0,1]$ denotes the confidence that pixel $i$ belongs to an authentic region. 
The detector is trained with pixel-wise supervision to separate natural null-space variations from tampering artifacts that violate the physical imaging constraints imposed by $\mathbf{A}_\phi$.

\subsection{Optimization}
We jointly optimize the entire pipeline, including the optical forward model $g_\phi$, the Nullspace Network $f_\theta$, and the tamper detector $d_\psi$, in an end-to-end fashion. 
This joint training allows the optics and neural modules to co-adapt for both faithful image reconstruction and reliable authenticity verification. 
The total objective combines reconstruction fidelity, perceptual consistency, and tamper discrimination:
\begin{equation}
\mathcal{L}_{\text{total}} = 
\mathcal{L}_{\text{rec}} + 
\beta\, \mathcal{L}_{\text{perc}} +
\lambda\, \mathcal{L}_{\text{cls}},
\end{equation}
where $\beta$ and $\lambda$ balance the relative importance of perceptual quality and tamper detection, respectively.

The reconstruction term enforces data fidelity ~$
\mathcal{L}_{\text{rec}} = ~\| \mathbf{x} - \mathbf{x}_p \|_2^2,
$
while the perceptual term encourages semantic similarity in a feature space extracted from a pretrained network $\mathcal{L}_{\text{perc}} = \sum_i \| \varphi_i(\mathbf{x}) - \varphi_i(\mathbf{x}_p) \|_2^2$, where $\varphi_i(\cdot)$ denotes the activation of the $i$-th layer in the reference network.
Finally, the classification loss supervises per-pixel authenticity prediction:
\begin{equation}
\mathcal{L}_{\text{cls}} = -[c \log \mathbf{m} + (1 - c) \log(1 - \mathbf{m})],
\end{equation}
with $c \in \{0,1\}^n$ denoting the ground-truth authenticity mask and $\mathbf{m} = d_\psi(\Pi_{\mathcal{N}}(\mathbf{x}_p))$ representing the predicted authenticity map.


By jointly optimizing these objectives, the system learns to (1) reconstruct images consistent with the optical forward model, (2) encode verifiable null-space signatures unique to the camera’s phase mask, and (3) detect deviations caused by digital tampering. 
This integrated training strategy naturally aligns the Nullspace Network’s reconstruction behavior with the tamper detector’s sensitivity, allowing the model to focus on physically plausible, authentication-relevant features.

Finally, because the physical phase mask $\phi$ and the learned models are co-optimized, the resulting system inherits an intrinsic unforgeability before the digital life cycle of the image: only images produced through the authentic optical path could produce consistent NOWAs and pass the verification stage. 
If the NSN were compromised, the absence of the optical encoding prevents an attacker from reproducing valid signatures, preserving the system’s physical integrity guarantees.



%% file: sec/4_results.tex
\section{Results}
\begin{table*}[t]
\centering
\setlength{\tabcolsep}{4pt}
\resizebox{1.0\textwidth}{!}{
\begin{tabular}{lcccccccccccccccc}
\toprule
\multirow{2}{*}{Method} &
\multicolumn{3}{c}{Stable Diffusion \cite{rombach2022high}} &
\multicolumn{3}{c}{ControlNet \cite{zhang2023adding}} &
\multicolumn{3}{c}{SDXL \cite{podell2023sdxl}} &
\multicolumn{3}{c}{RePaint \cite{lugmayr2022repaint}} &
\multicolumn{3}{c}{Lama \cite{suvorov2022resolution}} \\
\cmidrule(lr){2-4}
\cmidrule(lr){5-7}
\cmidrule(lr){8-10}
\cmidrule(lr){11-13}
\cmidrule(lr){14-16}
& F1 $\uparrow$ & AUC$\uparrow$ & IoU$\uparrow$ 
& F1 $\uparrow$ & AUC$\uparrow$ & IoU$\uparrow$ 
& F1 $\uparrow$ & AUC$\uparrow$ & IoU$\uparrow$  
& F1 $\uparrow$ & AUC$\uparrow$ & IoU$\uparrow$ 
& F1 $\uparrow$ & AUC$\uparrow$ & IoU$\uparrow$  \\
\midrule
MVSS-Net \cite{dong2022mvss}              & 0.178 & 0.488 & 0.103 & 0.178 & 0.492 & 0.103 & 0.037 & 0.503 & 0.028 & 0.104 & 0.546 & 0.082 & 0.024 & 0.505 & 0.022 \\
OSN \cite{wu2022robust}                     & 0.174 & 0.486 & 0.101 & 0.191 & 0.644 & 0.110 & 0.200 & 0.755 & 0.118 & 0.183 & 0.644 & 0.105 & 0.170 & 0.430 & 0.099 \\
PSCC-Net \cite{liu2022pscc}               & 0.166 & 0.501 & 0.112 & 0.177 & 0.565 & 0.116 & 0.189 & 0.704 & 0.115 & 0.140 & 0.469 & 0.109 & 0.132 & 0.329 & 0.104 \\
IML-VIT \cite{ma2023iml}              & 0.213 & 0.596 & 0.135 & 0.200 & 0.576 & 0.128 & 0.221 & 0.603 & 0.145 & 0.103 & 0.497 & 0.059 & 0.105 & 0.465 & 0.064 \\
HiFi-Net \cite{guo2023hierarchical}            & 0.547 & 0.734 & 0.128 & 0.542 & 0.735 & 0.123 & 0.633 & 0.828 & 0.261 & 0.681 & 0.896 & 0.339 & 0.483 & 0.721 & 0.029 \\
EditGuard \cite{zhang2024editguard} & 0.966 & 0.971 & 0.936 & 0.968 & 0.987 & 0.940 & \textbf{0.965} & 0.989 & \textbf{0.936} & 0.967 & 0.977 & 0.938 & \textbf{0.965 }& 0.969 & \textbf{0.934} \\
\textbf{NOWA}          & \textbf{0.993} &\textbf{ 0.999} &\textbf{ 0.987} & \textbf{0.992} & \textbf{0.999} & \textbf{0.985} & 0.929 & \textbf{0.997} & 0.867 & \textbf{0.974} & \textbf{0.999} & \textbf{0.949} & \textbf{0.965} & \textbf{0.999} & 0.933 \\
\bottomrule
\end{tabular}}
\caption{Comparison with other competitive tamper localization methods under different AIGC-based editing methods.}
\label{tab:agec_comp}
\end{table*}
\begin{figure*}[t!]
    \centering
    \includegraphics[width=\linewidth, height=7cm]{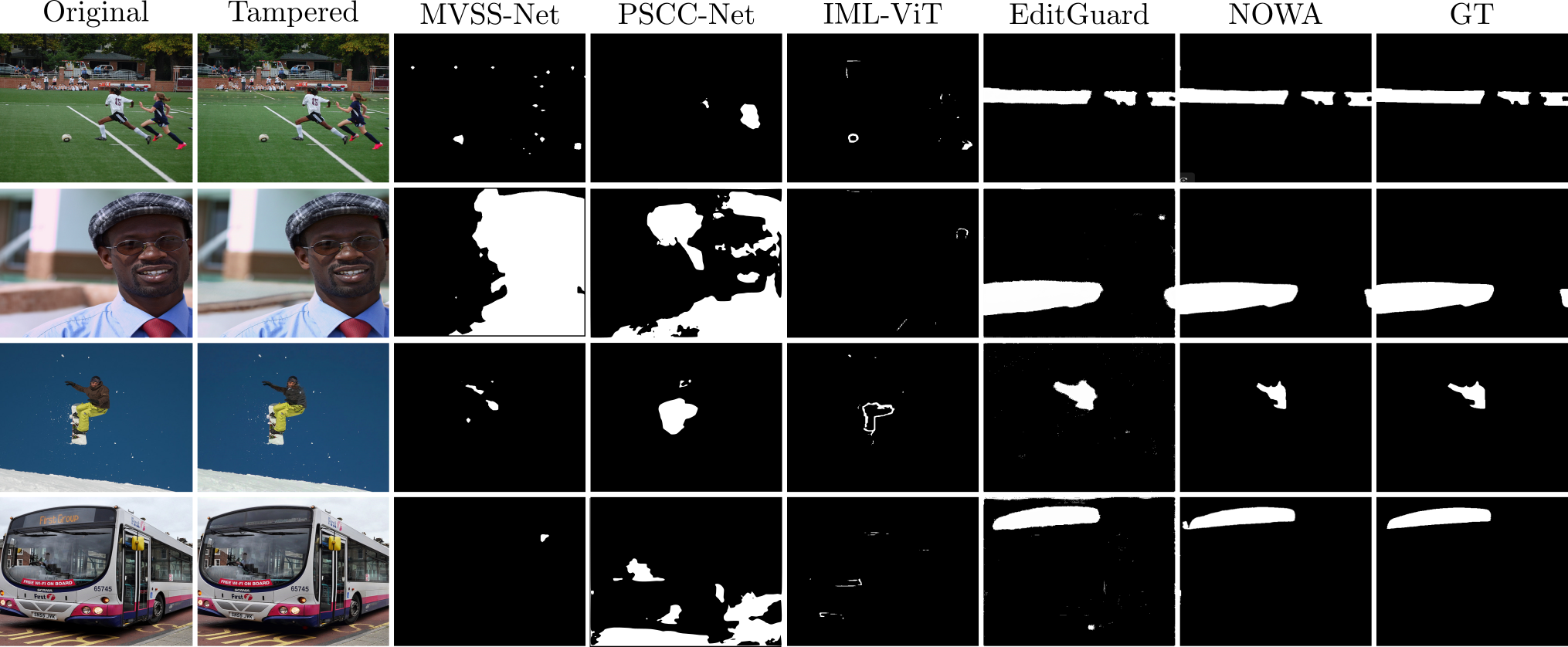}
    \caption{
        Qualitative comparison between the proposed method and state-of-the-art approaches. Our method produces more precise localization of manipulated regions and better preserves structural details compared to existing techniques.
    }
    \label{fig:sota_comparison}
\end{figure*}
\subsection{Experimental Setup}
To train our model, we utilize the FFHQ dataset~\cite{karras2019style}, which offers a diverse collection of high-resolution facial images, enabling robust evaluation of image authenticity and degradation resistance. For image modification, we adopt BiSeNetV2~\cite{yu2021bisenet} as the face parsing model to generate precise segmentation maps $\mathbf{m}$, followed by Stable Diffusion Inpainting~\cite{rombach2022high} to produce realistic and context-aware edits. Model evaluation is conducted on the EditGuard test set~\cite{zhang2024editguard}, comprising 1,000 images sampled from the COCO 2017 dataset~\cite{lin2014microsoft}. All experiments are performed on a single NVIDIA H100 GPU with 80 GB of VRAM. The size of the designed phase mask is $2.835$ mm $\times 2.835$ mm with a resolution of $256 \times 256$ pixels. We employ the AdamW optimizer with a learning rate of $1\times10^{-4}$, weight decay of $1\times10^{-2}$, and a batch size of 32. Each model is trained for $N_{\text{epochs}}$ epochs. To enhance generalization and robustness to real-world conditions, we apply data augmentations, including random cropping and additive Gaussian noise.

\subsection{Comparison to competing baselines}

We compare our method with representative models for manipulations localization and image authenticity detection. MVSS-Net \cite{dong2022mvss} leverages multi-view self-supervision to detect splicing and forgery cues from local texture inconsistencies. OSN \cite{wu2022robust} introduces object-centric supervision to improve robustness against editing operations. PSCC-Net \cite{liu2022pscc} employs progressive semantic context consistency for detecting subtle manipulations in complex scenes. IML-ViT \cite{ma2023iml} incorporates transformer-based long-range dependencies for improved pixel-level localization. HiFi-Net \cite{guo2023hierarchical} adopts a hierarchical multi-scale strategy to balance detail preservation and detection sensitivity. Finally, EditGuard \cite{zhang2024editguard} learns tamper localization, using diffusion-model-guided consistency maps to achieve high accuracy on AIGC-based edits. All baselines are evaluated using their official implementations and recommended configurations in five representative generative editing processes: Stable Diffusion \cite{rombach2022high}, ControlNet \cite{zhang2023adding}, SDXL \cite{podell2023sdxl}, RePaint \cite{lugmayr2022repaint}, and Lama \cite{suvorov2022resolution} to ensure a fair and comprehensive comparison.

As shown in Table~\ref{tab:agec_comp}, our hybrid optical–digital framework consistently outperforms all competing methods across all evaluation metrics. While prior approaches such as \textit{EditGuard} achieve strong performance (F1=0.97), our method further improves to an F1 of \textbf{0.993}, AUC of \textbf{0.999}, and IoU of \textbf{0.987}. This improvement highlights the benefit of embedding physically grounded NOWA at capture time. 
In contrast, purely digital detectors, such as MVSS-Net, OSN, or PSCC-Net, show degraded performance (F1$<$0.2) under generative editing, indicating limited resistance to content preservation manipulations typical of AIGC models. Even transformer and diffusion guided methods struggle to generalize to unseen edits, whereas our null-space–based detection remains stable and precise across all scenarios. 
Notably, our model was trained on FFHQ and evaluated on COCO, demonstrating strong cross-domain generalization. 


\begin{figure}[bt!]
    \centering
    \includegraphics[width=\linewidth]{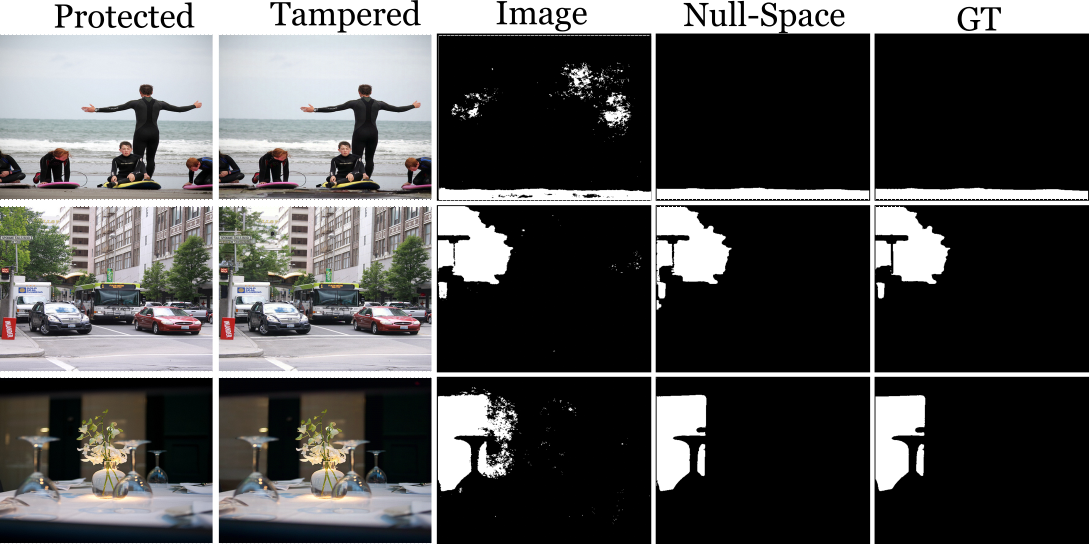}
    \caption{
    \textbf{Qualitative ablation of detector input.}
    Comparison of detected tamper when input to $d_\psi$ is the image or the null space. Using the image domain shows scattered false positives, while detector output on the null-space projection $\Pi_{\mathcal N}(\mathbf{x}_p)$ domain, 
    producing precise tamper localization.
    }\vspace{-2em}
    \label{fig:detector_ablation}
\end{figure}
\begin{figure*}[ht!]
    \centering
    \includegraphics[width=\linewidth]{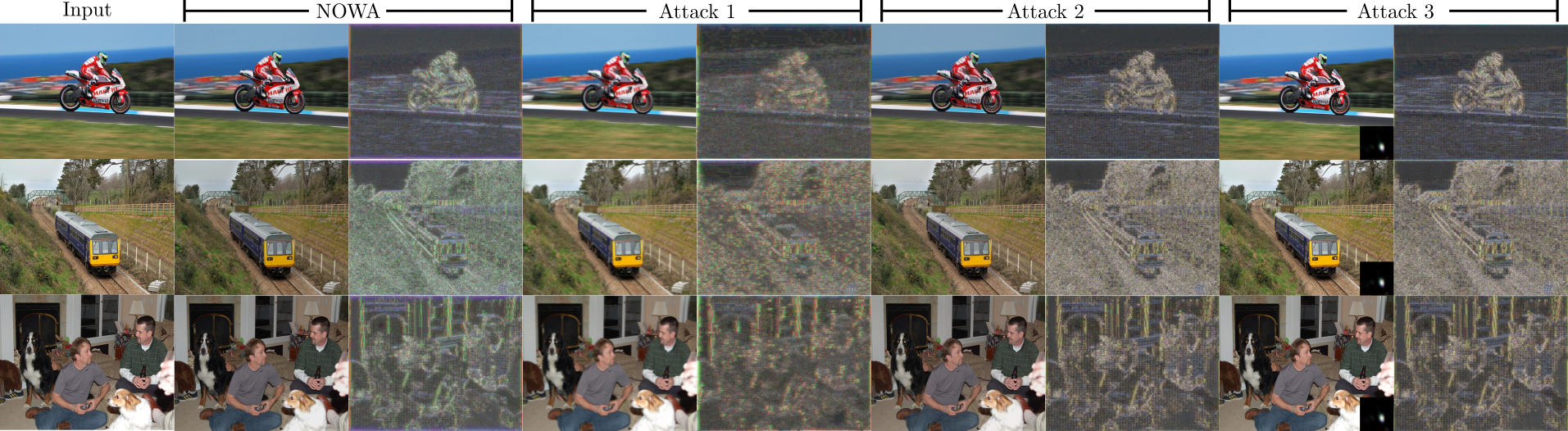}
    \caption{
   \textbf{Robustness of the proposed system against generative and analytical adversaries.} For each attack, we show the protected image produced by our system ($f_{\theta}$ output) alongside its corresponding \textit{Null-Space} projection. 
    Although counterfeit images closely replicate the visual appearance, they fail to reproduce the underlying null-space signature, enabling reliable detection and rejection by our method.
    }
    \label{fig:attack_results}
\end{figure*}
\subsection{Ablations}
\paragraph{Effect of null-space projection on detection performance.}
\label{sec:ablation_nullspace}


To evaluate the role of the null-space projection in our detection framework,
we compare two inputs to the classifier $d_\psi$:
(i) the protected image $\mathbf{x}_p$, and
(ii) its null-space projection $\Pi_{\mathcal N}(\mathbf{x}_p)$, which isolates the embedded physical signature.
Both settings use identical architectures and training protocols.

Using $\mathbf{x}_p$ directly yields a lower performance
(F1$\sim$0.75, AUC$\sim$0.82, IoU$\sim$0.69), suggesting that tampering leaves weak cues in the image domain.
In contrast, feeding the null-space projection produces near-perfect scores, demonstrating that
$\Pi_{\mathcal{N}}$ effectively suppresses natural image variation and exposes the invariant watermark.
Qualitative results in Fig.~\ref{fig:detector_ablation} further show that image-based predictions are noisy with scattered false positives,
whereas null-space projections enable clean and spatially coherent detection maps.
\vspace{-2em}
\begin{table}[]
\centering
\resizebox{\columnwidth}{!}{%
\begin{tabular}{cccccccc}
\hline
\multirow{2}{*}{\textbf{Method}} & \multirow{2}{*}{\textbf{Metrics}} & \multicolumn{2}{c}{\textbf{Gaussian Noise}} &  & \multicolumn{3}{c}{\textbf{JPEG}} \\ \cline{3-4} \cline{6-8} 
                           &     & $\mathbf{\sigma=1}$ & $\mathbf{\sigma=5}$ &  & $\mathbf{Q=70}$ & $\mathbf{Q=80}$& $\mathbf{Q=90}$ \\ \hline
\multirow{2}{*}{HiFi-Net \cite{guo2023hierarchical}}  & F1 $\uparrow$  &  0.236   & 0.156    &  &  0.315    &  0.404    &  0.476    \\
                                                     & AUC $\uparrow$ &  0.604   & 0.568    &  &  0.551    &  0.602    &  0.661    \\ \hline
\multirow{2}{*}{EditGuard \cite{zhang2024editguard}} & F1 $\uparrow$ &  0.908   &  0.871   &  &   0.552   &  0.813    &  0.785    \\
                           & AUC $\uparrow$ &  0.985   &  0.977   &  &   0.875   &   0.973   &  0.961    \\ \hline
\multirow{2}{*}{NOWA}      & F1 $\uparrow$  & 0.984    &  0.982   &  &   0.885   & 0.893     &  0.890    \\
                           & AUC $\uparrow$& \textbf{0.999 }   & \textbf{0.999}    &  &  \textbf{0.994}    & \textbf{0.994 }    &  \textbf{0.992}    \\ \hline
\end{tabular}%
}
\caption{Comparison of tampering localization under different levels of noise and compression degradations.}
\label{tab:degradations}
\end{table}

\paragraph{Effect of Degradations.}

As shown in Tab.~\ref{tab:degradations}, we evaluate robustness against common degradations (JPEG, noise) and advanced tampering (Stable Diffusion Inpaint~\cite{rombach2022high}). Our NOWA approach retains high localization accuracy with F1 scores around 0.9 even under compression, and minimal performance loss under Gaussian noise degradation. In contrast, the purely digital HiFi-Net~\cite{guo2023hierarchical} degrades significantly. This confirms that embedding the watermark at the optical level preserves the null-space signature and suppresses irrelevant variability, ensuring reliable detection even under severe perturbations.
\vspace{-1em}
\paragraph{Effect of Learned Phase Mask. }
We also examine a configuration without the phase mask. The optical system still has a null space, allowing the NSN to imprint a NOWA; however, this signature is significantly less structured than the one produced with the learned mask. As a result, the reconstructed images retain only a faint watermark, and the detection module struggles to distinguish genuine deviations from natural variation (F1=0.89, IoU=0.81). These observations confirm that the proper design of the null space is essential for making the NOWA strong, stable, and discriminative. 
\begin{figure*}[ht!]
    \centering
    \includegraphics[width=\linewidth]{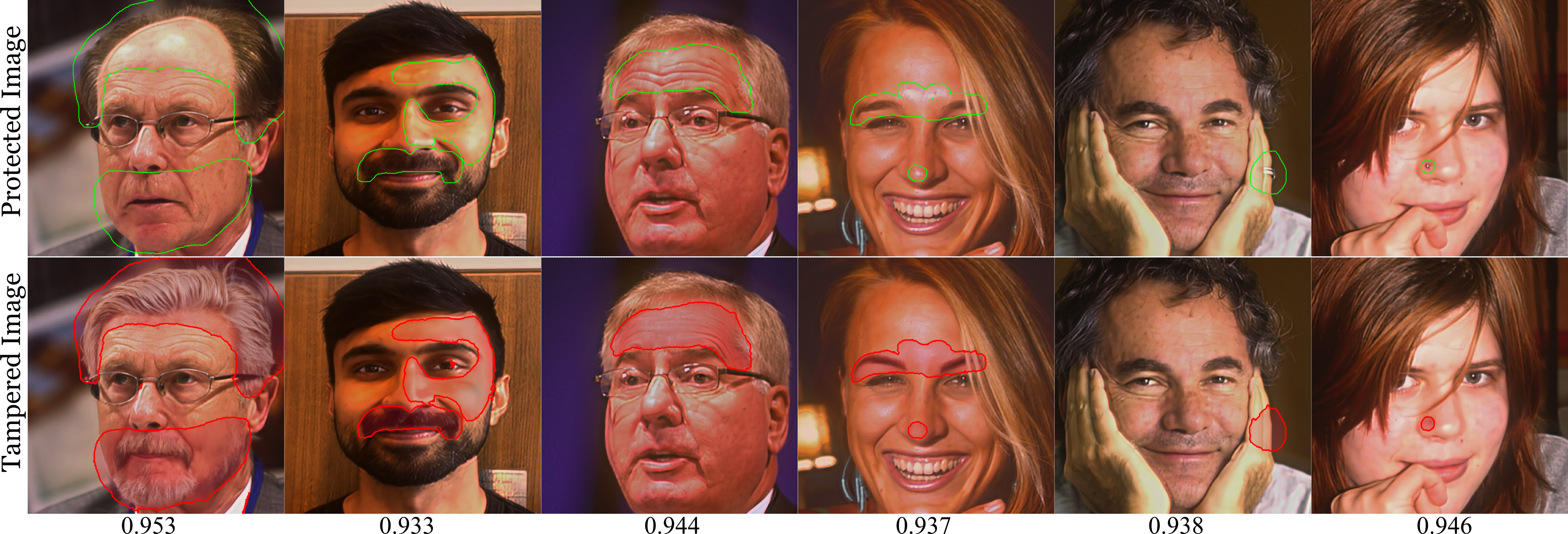}
    \caption{\textbf{Tamper localization from real captures.}
    Each column shows (top) the protected image with the target edit region outlined in green, and (bottom) the tampered image with the estimated manipulation mask overlaid in translucent red. Edits were manually made using Photoshop tools. The IoU scores below each example indicate strong localization accuracy and robust detection of real digital edits, as supported by the visual results. 
    }
    \label{fig:real_results}
\end{figure*}
\subsection{Adversarial Robustness Evaluation}

We assess robustness under three adversarial scenarios in which the attacker attempt to replicate the distribution of protected images having access to $7{,}000$ camera outputs $\mathbf{y}$ or protected reconstructions $\mathbf{x}_p$, along with $10{,}000$ unprotected public images.
The adversarial attacks are as follows:
\vspace{-2em}
\paragraph{Attack 1: Camera Imitation.}
From camera outputs $\mathbf{y}$, the adversary attempts to replicate the behavior of the physical camera by learning a generative model $\mathcal{G}_1$ that maps unprotected images to camera outputs $
    \mathcal{G}_1 : \mathcal{X}_{\text{u}} \rightarrow \mathcal{Y}'.$
The resulting fake camera images $\mathbf{y}' = \mathcal{G}_1(\mathbf{x}_{\text{u}})$ are then passed through the original protection network $f_\theta$ (asumming is exposed) to generate fake protected images $\mathbf{x}_p'$. 
\vspace{-1.5em}
\paragraph{Attack 2: Protected Image Imitation.}
In this case, the adversary directly targets the distribution of protected images by training a generative model $\mathcal{G}_2$ that learns to map public unprotected data into the protected domain $\mathcal{G}_2: \mathcal{X}_{\text{u}} \rightarrow \mathcal{X}'_{\text{p}}$. 
The generated images $\mathbf{x}'_p = \mathcal{G}_2(\mathbf{x}_{\text{u}})$ are subsequently processed by the image verification network $d_\psi$. 
\textbf{Attack 3: Blind Deconvolution.}
From camera outputs, the adversary can also attempt to estimate the system PSF  via blind deconvolution techniques and apply it to arbitrary public images 
$
\hat{\mathbf{y}} = \hat{\mathbf{A}}_\phi \mathbf{x}_{\text{u}}.
$
The generated $\hat{\mathbf{y}}$ is then passed through the protection function $f_\theta$ (asumming is exposed) and the verification module $d_\psi$. 

For $\mathcal{G}_1$ and $\mathcal{G}_2$, we employ a diffusion-based image-to-image translation model~\cite{parmar2024one}, and for blind deconvolution we use the method of Eboli \textit{et al.}~\cite{eboli2022fast}. 
As shown in Fig.~\ref{fig:attack_results}, all three attacks produce images that are visually similar to our system; however, they fail to reproduce the intrinsic null-space correlations imposed by the physical camera model. 
Specifically, for $1,000$ spoofed images per attack, our system achieves detection accuracies (F1 scores) of 0.901, 0.913, and 0.946 for attacks 1, 2, and 3, respectively,
confirming the robustness of our hybrid optical–digital design against both learned and physics-based adversaries. 
Notably, even when the NSN is accessible to the attacker (Attacks~1 and~3), the absence of the authentic optical system prevents the generation of valid NOWAs.
Likewise, the failure of Attack~2 shows that, without the optical encoding, the protected and natural image distributions remain statistically indistinguishable. 
\vspace{-0.8em}
\section{Real Captures Experiments}
We verify the proposed framework in a real-world capture experiment by translating the forward model $\mathbf{A}_\phi$ used in simulation to a physical optical prototype. The setup comprises a Canon EOS5DMarkIV DSLR coupled with a 50mm f/1.8 YONGNUO lens and the designed phase mask fabricated using two-photon lithography on a fused-silica substrate. The phase mask was placed in the aperture plane of the YONGNUO lens. 
Additional fabrication and optical setup details are provided in the Supplementary Material. For accurate protection under real imaging conditions, we fine-tune the Null-Space Network $f_\theta$ and the detector $d_\psi$ for 30 epochs with a learning rate of $1\times10^{-5}$, using scenes from the FFHQ dataset rendered through the measured point spread functions (PSFs) of the real prototype. Captured images, fine-tuning details, and calibration, including PSF measurement and alignment procedures, are provided in the Supplementary Material.

To simulate realistic tampering, we manually edited captured protected images using Adobe Photoshop, employing the lasso selection tool followed by generative fill prompts based on Adobe generative AI technology. These edits introduce semantically plausible modifications but measurement-inconsistent modifications. 
Despite the realism of the generated content, our detector accurately localized the tampered regions with high precision, achieving pixel-level consistency across multiple test captures. Figure~\ref{fig:real_results} shows real-world examples of protected, edited, and detected tampering demonstrating the robustness of the proposed NOWA to authentic image manipulations in real-world captures. 


%% file: sec/5_conclusion.tex
%
\vspace{-1em}
\section{Discussion}
Our results demonstrate that jointly optimizing optical and digital components enables a physically grounded authentication system. By embedding optical cues at capture through the phase mask, the Null-space Optical Watermark (NOWA) becomes a hardware-dependent signature, invisible in the image but recoverable via the null-space projection. Furthermore the NOWA acts as a non-invertible physical key, creating a security asymmetry, valid signatures cannot be reproduced without the genuine phase mask, even if the digital pipeline is exposed. 
Limitations include the small and fixed aperture size of the phase mask, which constrains light throughput. Further, for larger apertures the depth-dependent behavior of the PSF should be accounted for correct computation of the null-space. Additionally, as observed in the real capture experiments, optical misalignment or imperfect calibration can degrade null-space estimation and violate measurement consistency. 
Finally, we only assume attack models that have protected $\mathbf{x}_p$ or captured images $\mathbf{y}$. Future work will investigate randomized or key-based digital embeddings to mitigate potential attacks that exploit access to large paired $(\mathbf{x}_p, \mathbf{y})$ datasets for approximating the imaging operator.

\section*{Acknowledgments}
This work was supported by NSF award IIS-2107313
and DoD ARO W911NF-24-2-0008. We thank Aniket Dashpute and Vivek Boominathan for helpful discussions. We thank Salman Khan for fabricating the phase mask.

%% file: sec/X_suppl.tex
\clearpage
\setcounter{page}{1}
\maketitlesupplementary

\section{Hardware Prototype and Fabrication}
\label{sec:fabrication}

In this section, we provide the specific manufacturing parameters and assembly details for the physical prototype used in our real-world validation.

\subsection{Phase Mask Fabrication}
The height profile $h_\phi$ of the phase mask was fabricated via Two-Photon Polymerization 3D lithography on a (10mm $\times$ 10mm) fused-silica substrate. We utilized a Nanoscribe Photonic Professional GT laser writing system operating in Dip-in Liquid Lithography (DiLL) mode. For reliable printing, the height map of the
designed phase mask was discretized into 6 layers with steps of 200 nm. We employed the IP-DIP photoresist, which has a refractive index of $n=1.52$ at $550$ nm, closely matching the fused-silica substrate to ensure structural stability. A $63\times$ objective was used to focus the laser into the photoresist. 

\subsection{Optical Assembly}
The optical setup consists of a Canon EOS 5D Mark IV DSLR (Full-frame sensor, $6.5\,\mu\text{m}$ pixel pitch) and a commercial YONGNUO $50\text{mm}$ f/1.8 lens. 
To integrate the phase mask into the system, the lens was disassembled to access the physical aperture stop (see Fig.~\ref{fig:aperture} left).
Then, the fabricated phase mask was cut to a smaller size to fit the camera's aperture and attached to it using carbon tape. The substrate was cut to a smaller size to fit into the camera's aperture (see Fig.~\ref{fig:aperture} right). 

\begin{figure}[htbp]
  \centering
    \centering
    \includegraphics[width=\linewidth]{figures/lens.png}
    \caption{\textbf{Optical encoding}.  A commercial YONGNUO lens is disassembled to attach the design phase mask on the back side of the lens aperture. The inset shows the quantized designed phase mask.}
  \label{fig:aperture}
\end{figure}

\section{PSF Measurement}
\label{sec:calibration}

\begin{figure}[htbp]
  \centering
    \includegraphics[width=\linewidth]{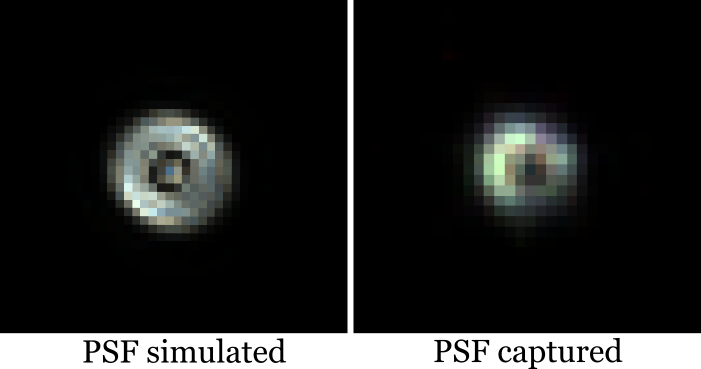}
  \caption{\textbf{PSF Calibration.} Comparison between the optimized simulated PSF (left) and the experimentally measured PSF from the physical prototype (right). The similarity in structure confirms the fidelity of the fabrication process.}
  \label{fig:psf_comparison}
\end{figure}

A critical step in bridging the simulation-to-reality gap is the accurate characterization of the physical Point Spread Function (PSF) generated by the fabricated mask.
We measured the system PSF using a point source setup:
1. A $20\,\mu\text{m}$ pinhole was back-illuminated by a broadband white LED source.
2. The prototype camera was aligned to the optical axis.
3. The distance between the pinhole and camera is 50 cm.
Figure~\ref{fig:psf_comparison} demonstrates the agreement between the simulated PSF and the physically measured PSF. The observed deviations are attributed to fabrication tolerances and misalignment of the mask relative to the optical axis.

\section{Fine-Tuning and Implementation Details}
\label{sec:finetuning}

While the Null-Space Network $f_\theta$ and detector $d_\psi$ were initially trained in a purely simulated environment, real-world optical aberrations and sensor noise introduce a domain shift. To ensure robust protection in physical experiments, we performed a fine-tuning stage. Ideally, one would collect thousands of real-world captures to fine-tune the model; however, acquiring such a large-scale paired dataset imposes a prohibitive workload.

To address this, we adopted a hybrid training strategy. Instead of relying on the analytical wave-optics model, we employ the \textit{measured} PSFs described in Section~\ref{sec:calibration} combined with clean images from the FFHQ dataset to synthetically generate optically encoded measurements $\mathbf{y}$. We further augmented these measurements by adding read noise and photon shot noise calibrated to the Canon 5D sensor characteristics (ISO 100). With these measurements, we fine-tune the Null-Space Network $f_\theta$ and the detector $d_\psi$ for 30 epochs with a learning rate of $1\times10^{-5}$.



\section{Real-World Capture Procedure and Additional Analysis}
\label{sec:results}

We conducted the capture experiments in a controlled, realistic indoor lighting scenario. We displayed all test scenes on a high-resolution LCD monitor rather than using printed targets. This approach ensured high spatial fidelity, stable illumination, and rapid switching between scenes. The monitor was positioned perpendicular to the optical axis at a fixed distance of 50 cm from the camera aperture. 
All captures were made using the Canon EOS-5D Mark IV with the 50 mm f/1.8 YONGNUO lens and the fabricated phase mask mounted at the aperture stop. All images were captured in RAW format to avoid in-camera processing. Captured images are cropped and resized to 512$\times$512, matching the resolution used in our simulations.

\begin{figure*}[ht!]
  \centering
    \includegraphics[width=1\linewidth]{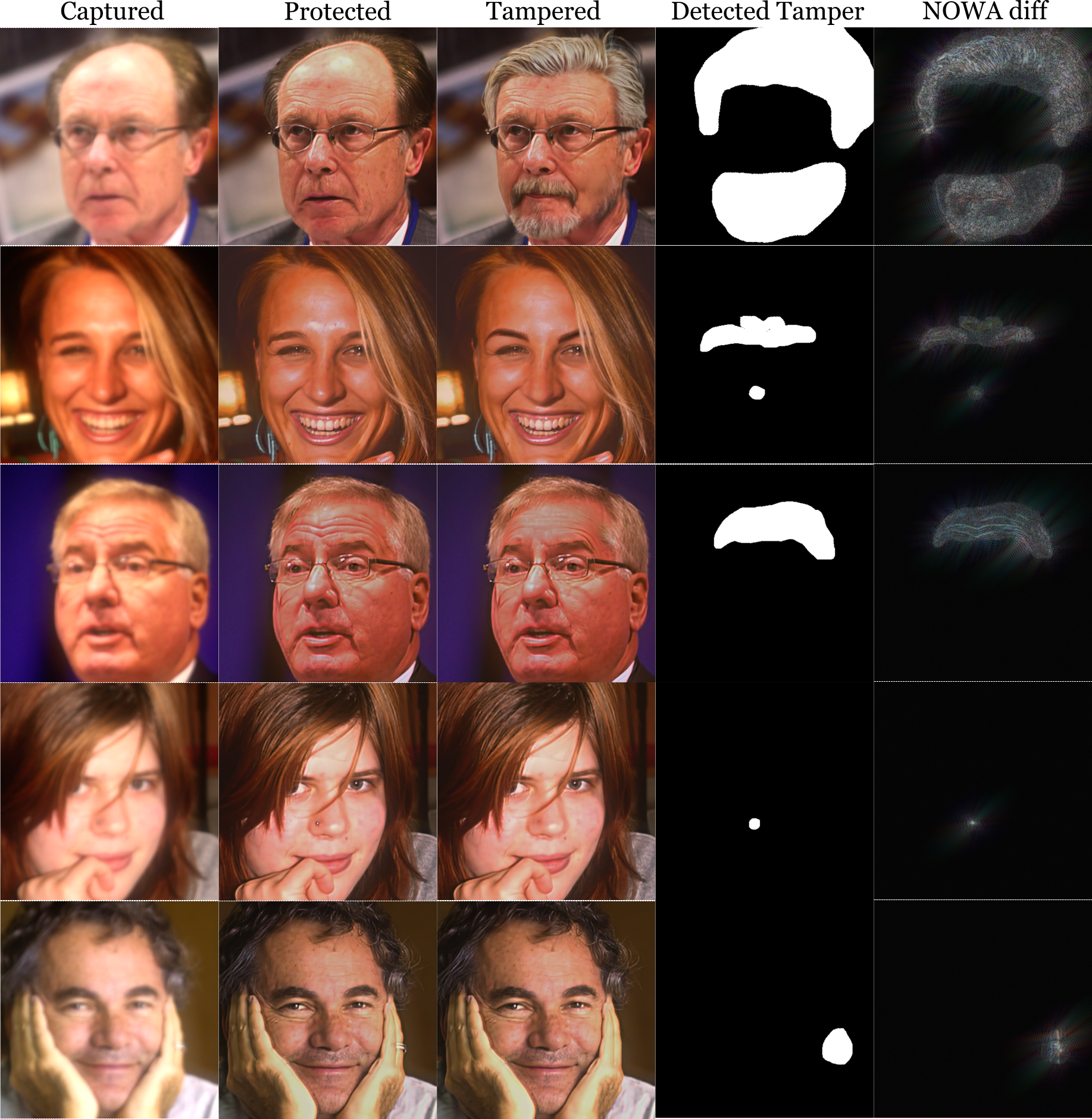}
  \caption{ \textbf{Real-world prototype results.} From left to right: raw capture $\mathbf{y}$, protected reconstruction $\mathbf{x}_p$, tampered image (Photoshop Generative Fill), detected tamper mask, and NOWA discrepancy between authentic and edited captures. Manipulations introduce measurement-inconsistent NOWA residuals that are reliably localized by our detector.}
  \label{fig:real_captures}
\end{figure*}

The first column of Fig.~\ref{fig:real_captures} shows the raw measurements $\mathbf{y}$ captured with the prototype. Passing these measurements through the fine-tuned NSN $f_\theta$ yields the protected reconstructions $\mathbf{x}_p$, shown in the second column. We can recover the scene content with high visual fidelity. To simulate realistic manipulations, we manually edited the protected images using Photoshop (lasso selection + Generative Fill). The resulting tampered images and their corresponding detection masks produced by our NOWA framework are presented in the third and fourth columns, respectively. These results confirm that the robustness observed in simulation carries over to real-world captures, demonstrating the physical-layer security provided by the NOWA. The final column visualizes the discrepancy between the NOWA of the authentic and edited images. High-frequency deviations emerge within the manipulated regions, while mild low-frequency differences appear near the edges. Still, our proposed detector reliably isolates the tampered areas.

\paragraph{Discussion and Handling of Depth Variations.}

Although our real-capture experiments use a high-resolution LCD monitor positioned at a fixed distance, this setup faithfully represents the optical behavior expected in real scenes for two reasons. First, the phase-mask prototype operates with a small effective aperture, producing a naturally large depth of field. In this regime, objects lying within a broad depth range produce sensor measurements whose point spread function (PSF) and corresponding null-space structure vary only minimally with respect to depth. Second, bsuooecause the NOWA embedding arises from the null space of the imaging operator $\mathbf{A}_\phi$, and because this operator is largely invariant within the depth-of-field limits, the NOWA behaves consistently whether the content is displayed on a monitor or originates from a real 3D environment.

For scenes with large depth variations or macro-scale imaging, the PSF becomes depth-variant. In such cases, one can extend the framework by calibrating a small set of depth-dependent PSFs or by incorporating a depth-conditioned NSN stage. Both are compatible with our current formulation and require the proper augmentations to the reconstruction pipeline. We leave the development of depth-aware calibration strategies as an interesting future direction, enabling the NOWA framework to handle scenes beyond the current depth-of-field regime.

\begin{figure*}[ht!]
  \centering
    \includegraphics[width=1\linewidth]{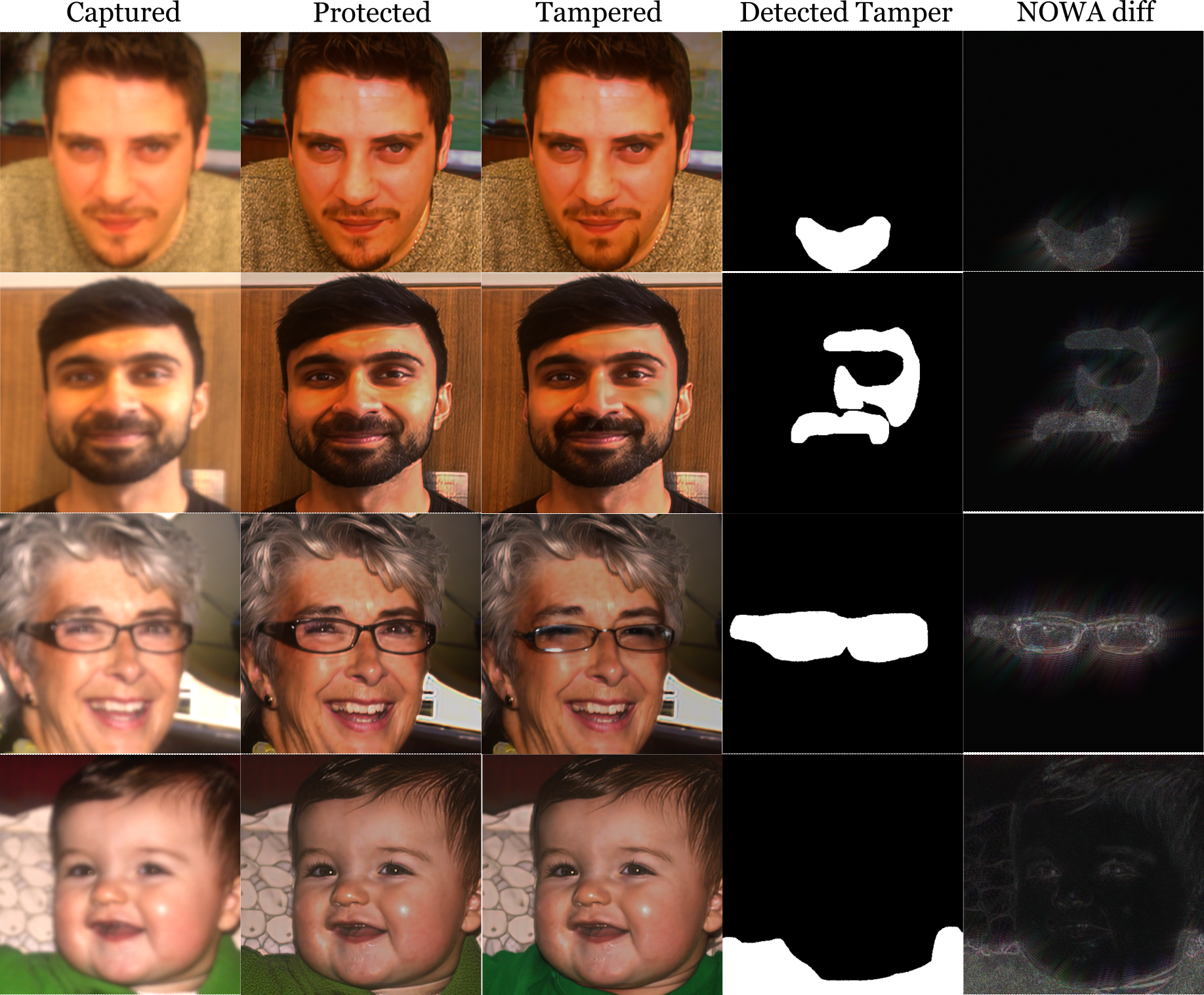}
  \caption*{ Figure 8. (cont.). \textbf{Real-world prototype results.} From left to right: raw capture $\mathbf{y}$, protected reconstruction $\mathbf{x}_p$, tampered image (Photoshop Generative Fill), detected tamper mask, and NOWA discrepancy between authentic and edited captures. Manipulations introduce measurement-inconsistent NOWA residuals that are reliably localized by our detector.}
\end{figure*}

\color{black}
\paragraph{Robustness against real-world camera imperfections } 

The phase mask induces a deterministic, structured PSF, whereas contaminants (e.g., dust) and minor aberrations introduce stochastic optical distortions. The null-space projection acts as a matched filter, isolating the watermark's structured residuals from uncorrelated noise. However, in our real-world experiments, imperfect calibration or slight misalignment  degrade null-space estimation that raises the energy baseline of the null space. Even after precise PSF calibration, the residual null-space energy $\|\Pi_\mathcal{N}(\hat{\mathbf{x}}_r)\|$ increases from a baseline of $\sim 10^{-5}$ (not 0  due to numerical precision limits) in simulation to $\sim 10^{-4}$ in real-world captures. While this indicates minor spectral leakage of unwanted signals, this ``real-world noise floor" remains significantly lower than the residuals induced by tampering. Consequently, the detector successfully differentiates between the stochastic artifacts of physical imperfections and the specific null-space violations of a forgery.